\documentstyle[12pt]{article}
\begin{document}

\renewcommand{\Large}{\large}
\renewcommand{\huge}{\large}

\begin{titlepage}
\baselineskip .15in
\begin{flushright}
WU-AP/33/93
\end{flushright}

{}~\\

\vskip 1.5cm
\begin{center}
{\bf
\vskip 1.5cm
{\large\bf Stability of Non-Abelian Black Holes\\[.5em]
 and Catastrophe Theory}

}\vskip .8in

{\sc K. Maeda}$^{(a)}$, {\sc T. Tachizawa}, {\sc T. Torii}
\\[.5em]
{\em Department of Physics, Waseda University,
 Tokyo 169-50, Japan}\\[1em]
and\\[1em]
{\sc T. Maki} \\[.5em]
{\em Department of Physics, Tokyo Metropolitan University,
Tokyo 192-03, Japan}
\end{center}
\vfill
\begin{abstract}
Two types of self-gravitating particle solutions found in
several theories
with non-Abelian fields are smoothly connected by a family
of
non-trivial black holes.  There exists a maximum point of
the black
hole entropy, where the stability of solutions changes.
This criterion is
universal, and the changes in stability follow from a
catastrophe-theoretic
analysis of the potential function defined by black hole
entropy.
\end{abstract}


\vfill
\begin{center}
October, 1993
\end{center}
\vfill
(a)~~electronic mail: maeda@jpnwas00.bitnet ~or~
maeda@cfi.waseda.ac.jp\\
\end{titlepage}

\normalsize
\baselineskip = 22pt

After  Bartnik and McKinnon discovered a non-trivial
particle-like structure (BM
particle) in the Einstein-Yang-Mills theory\cite{BM}, a
variety of
self-gravitating structures with non-Abelian fields have
been found. Besides the
BM particle, researches have discovered the  colored black
hole\cite{Bizon}, the
Skyrmion\cite{skyrmion,DHS} or the Skyrme black hole\
cite{LM,DHS,TM}, the
monopole\cite{monopole,LNW,BFM} or the black hole in
monopole (monopole black
hole)\cite{LNW,BFM,AB}, the  particle solution with
massive Proca field (Procaon) or
the Proca black hole\cite{GMO}, and the sphaleron\
cite{sphaleron,GMO} or
sphaleron black holes\cite{GMO}.

One of the most important questions about these self-
gravitating
non-Abelian structures is, are they stable?  The BM
particle  and the colored
black hole are unstable against  radial perturbations,
while both the Skyrmion
and  the monopole, and the corresponding black hole
solutions, are stable. The
sphaleron and its black hole solution may be unstable
because of their
topological structure. Are  there any common properties in
those non-Abelian
structures?  Can we find any universal understanding for
them?  Answering these
questions is the main purpose of the present paper. We
will soon show that there
is a universal picture for self-gravitating non-Abelian
structures that
incorporates these black hole solutions, and that accounts
for their stability
properties via a catastrophe-theoretic analysis of the
black hole entropy, $S$
(=the area of event horizon/4) regarded as a potential
function.

We have re-analyzed 5 models, which  are listed in Table
1.
Some known results concerning these models are also
summarized in the table.
Remarkably, except for the colored black hole and the
monopole black hole,
all solutions share the following properties\cite{note1}:
\\
{\bf (1)} There are two particle-like solutions.  One
corresponds to the known
particle solution without gravity\
cite{skyrmion,sphaleron}, and the
other has properties similar to those of the BM
particle\cite{BM,DHS,GMO}.
{\bf (2)} Two branches of black hole
solutions, which leave from two particles, bifurcate at
some
critical
horizon radius\cite{LM,GMO}.  Beyond this critical point,
where the
black hole has a maximum mass and a maximum entropy, there
exists no non-trivial
structure. The upper branch in Fig. 1 has
larger entropy than that of the lower branch. Hence, we
shall call
each of them high- and low-entropy branches, respectively.
The low-entropy branch is similar to the colored black
hole solution, and the high-entropy branch approaches the
Schwarzschild
black hole in the ``low energy" limit\cite{TM}. Here, the
``low energy" means
that the mass of the particle is much smaller than the
Planck mass $m_P \equiv
G^{-1/2}$.   It is realized in the limit as $\mu \
rightarrow 0$, where $\mu$ is
a mass of the relevant non-Aberian field, e.g.,  $\mu = g
\Phi_0$ (the vacuum expectation
value of the Higgs field) for the Einstein-Yang-Mills-
Higgs system, or $\mu =
g_Sf_S$ (two coupling constants of Skyrme field)
for the Eisntein-Skyrme
system. (Note that $g_S^2=4 \pi g^2$ in our notation).
On the other hand, in the limit of ``high energy", no
solution exists.
It disappears around $\mu \sim m_P/g$.
{\bf (3)} The
high-entropy branch is stable (except for the sphaleron
solution, but see later),
while the low-entropy branch is unstable\
cite{DHS,LM,SZ,HDS}.   {\bf (4)} The
specific heat in the high-entropy branch is always
negative, while the
specific heat in the
low-entropy branch changes its sign a few times\cite{TM}.

In order to obtain a universal picture with the properties
{\bf (1)} - {\bf (4)} above, we
have re-analyzed the 5 models listed in Table 1 and found
the following new
results\cite{TMMT2}:
{\bf (5)} Fixing the horizon
radius $r_H$, there are two black hole solutions with
different masses.
Those
two branches are bifurcated at some critical radius.  In
the mass-radius
($M$-$r_H$) plane, the solution curve has a cusp at this
critical point $C$.
(see the figure 1). The stability changes at this cusp,
that is, the
high-entropy branch  is stable while the low-entropy one
is unstable
against radial perturbations. {\bf (6)} If we draw the
solution curve in the
three dimensional space of the mass $M$, the entropy of
the black hole $S$, and
the field strength at the horizon
$B_H \equiv ({\rm Tr} F^2)^{1/2}|_{horizon}$, it becomes
smooth. (see the figure
2). Here, the expression $B_H$ has been  used because only
the radial component
of magnetic part of non-Abelian field is finite at the
horizon.  Only the
projection onto the $M$-$S$ plane (and then  onto the
$M$-$r_H$ plane) provides
a cusp.   The cusp, at which the stability changes,
corresponds to a turning
point in the three dimensional picture, where the black
hole entropy takes
the maximum value.

The appearance of a cusp as a critical point of stability
is often
discussed in catastrophe theory\cite{Gibbons,Thompson} and
in its application to astrophysics
\cite{SKM,KKO}. In the present case, if we
regard $S$, $M$ and $B_H$ as a potential function, a
control parameter, and a generalized coordinate,
respectively, we may apply catastrophe theory to the
present stability problem
as follows.
In catastrophe theory, solutions are regarded as extremal
points
on the Whitney surface, $S=S(M, B_H)$, when the
control parameter $M$ is fixed. If a solution is
a maximal point, then that solution is stable because its
entropy is maximal. On
the other hand, if it is a minimal point, then it is
unstable. At the maximum
entropy, the solution turns out to be an inflection point,
beyond which
there is no extremal point, i.e., there is no black hole
solution\cite{TMMT2}.

We may wonder what happens with the sphaleron black hole,
because
both its high- and low-entropy branches are unstable for
topological reasons.
Is this consistent  with our
interpretation of stability via catastrophe theory?  When
we
discuss stability, in general there are many modes to be
investigated.  A general argument about the instability of
the sphaleron is based on a
topological analysis\cite{sphaleron}, which does not
choose any specific mode.  On the
other hand, when we discuss stability change using
catastrophe
theory, we focus on some specific mode.
For the sphaleron without gravity, the stability analysis
with a
spherically symmetric ansatz was done\cite{AKY}.
It was explicitly shown that there is only one unstable
mode. Although
no analysis has been made, so far, for the case of the
gravitating sphaleron or
the sphaleron black hole, we
guess that it may be stable in the high-entropy branch
against radial
perturbations except for one unstable mode corresponding
to the above.  In the low-entropy
branch, some of stable modes become unstable. The
sphaleron black hole
picks up at least one more unstable mode beyond the
critical point.
In this sense, we argue that the  high-entropy
branch is ``stable" while the low-entropy one is unstable.
If this is so, then catastrophe theory accounts correctly
for the stability of
black holes even in the sphaleronic case.

{}From the above discussion, we see that we can classify
non-Abelian black holes
into two types,{\bf (A)} and {\bf (B)}:\\
{\bf (A)} High-entropy ``neutral" types\\
The high-entropy branch is ``stable".
The field strength at the horizon($B_H$) is still small as
well as the
black hole is  globally neutral.  The black hole is
approximately neutral.
We may adopt the following picture for this type of black
hole.
The non-Abelian structure may be approximated as a uniform
vacuum energy density
$\rho_{vac}$ with a sphere whose radius is the Compton
wave
length of the massive non-Abelian field.  As for the black
hole solution,
the horizon must exist in the
region of uniform vacuum energy.
Otherwise, non-trivial non-Abelian structure is swallowed
by the
black hole,
resulting in a trivial Schwarzschild  solution.
This explaines why there is an upper bound on the mass or
horizon radius
for this non-trivial solution.
{}From our picture, the high-entropy ``neutral" black hole
near the horizon
is approximated by the
Schwarzschild-de Sitter spacetime with  the cosmological
constant=$8\pi G \rho_{vac}$.  In the limit of ``low
energy", the solution
approaches the Schwarzschild black hole.  The negative
specific heat is
also consistent with that of the Schwarzschild or
Schwarzschild-de Sitter
spacetime.
The self-gravitating particle approaches
the known particle solution in a Minkowski
background\cite{skyrmion,sphaleron}. Such a particle
can exist without gravity.   \\
{\bf (B)} Low-entropy ``locally charged" types\\
The low-entropy branch is unstable. The structure of this
type of black hole is
quite similar to the colored black hole.  Although the
black hole  is
globally neutral, $B_H$ does not vanish at the horizon.
Its
value is rather large.  An effective charge appears near
the black hole
horizon. Furthermore, in the  ``low energy" limit, the
solution
approaches the colored black hole\cite{DHS,TM}. The non-
trivial structure in
this case is due to the kinetic term of non-Abelian gauge
field,
${\rm Tr} F^2$. Gravity must play an essential role in
this non-trivial
structure, because the BM particle cannot
exist without gravity and the mass scale is about $m_{BM}
\sim m_{P} /g$,
which is almost  independent of $\mu$ (or $\Phi_0, f_S$).
Just as we found strange
behavior in the specific heat of a colored black hole\
cite{TM}, we find
a few times changes of its sign (see Table 1).

As for the excited state, i.e., higher-node solutions, we
find that a
similar cusp exists\cite{TM,TMMT2}.
We expect that when the solution goes beyond this cusp
(the maximum
entropy point), another instability will appear.  Since
the colored black
hole has $n$ unstable modes for an $n$-node solution\
cite{SZ,SW}, we expect
that the high-entropy branch of $n$-node solutions has
$(n-1)$ unstable modes
while the low-entropy branch has $n$ unstable modes.

We have, so far, discussed all known non-Abelian
structures except for the
monopole black hole and the colored black hole, and have
presented a universal
picture.   Although the colored black hole
does not always have all the properties we have discussed
above, it should be
included in our universal picture.  Because the colored
black hole and the
Schwarzschild black hole  are obtained  exactly as ``low
energy" limits of the
low- and high-entropy branches, respectively. The only
exceptional
solution is the monopole or the monopole black hole, which
is  globally charged.

It should be stressed, however,
that although the monopole black hole has different
properties from
the types {\bf (A)} and {\bf (B)} above and shows more
complicated
behaviours\cite{LNW,BFM,AB},  the catastrophe theory is
again applied to the
stability analysis\cite{TMMT}.
Depending on the parameters $g, \lambda$, and $\Phi_0$ in
the
Einstein-Yang-Mills-Higgs system,
there seem to be the following two cases\
cite{LNW,BFM,AB,TMMT}:\\
{\bf (I)} The mass of the monopole black hole increases
monotonically as
entropy increases and the solution eventually reaches at a
bifurcation
point $B$ with the RN black hole branch.  No cusp appears.
The monopole black hole is stable, while the
RN solution becomes unstable beyond this bifurcation point
$B$\cite{LNW}.\\
{\bf (II)} For some range of parameters, the solution
curve of
the monopole black holes has a cusp $C$ in the $M$-$S$
plane\cite{AB,TMMT}, where the black hole has the maximum
entropy. There are two
solutions with the same horizon radius (the same entropy)
but different masses
just  as with the other type of non-trivial black holes.
When the radius gets small in the second (low-entropy)
branch, the solution
either may merge to the RN black hole at a bifurcation
point $B$ or might
reach to another particle solution (BFM particle\
cite{BFM}).
We guess that the second low-entropy branch is
unstable while the first high-entropy branch is stable
(see \cite{AB}).  The RN
black hole is stable before the bifurcation point $B$, but
it becomes unstable beyond  $B$.

All these behaviors {\bf (I)} and {\bf (II)}, including
the stability of
RN black hole, follow easily
from  catastrophe theory, if the entropy is regarded as
the
potential function\cite{note2}. The entropy $S$ with a
fixed mass $M$ is maximal for
the stable branch but becomes minimal for the unstable
branch\cite{TMMT}.

In this letter, we have re-analyzed the known non-Abelian
black holes,
as well as  the corresponding
the self-gravitating particle-like  solutions.  We find  a
universal picture:
The globally neutral solutions  are classified into  two
types depending on
whether they are almost neutral or locally charged.  The
``neutral" type
(high-entropy branch) is similar to the Schwarzschild-de
Sitter solution  and
``stable" (see the previous discussion for sphaleron black
holes).  The
``locally charged" type (low-entropy branch) is like the
colored black hole
 and unstable. Its specific
heat changes the sign a few times with respect to the
mass.  When those two
types coincide, the entropy becomes maximum. Catastrophe
theory can be
applied to analize the stability of these black holes. One
stable mode
in the high-entropy branch becomes unstable beyond the
bifurcation point.
This may be true also for the sphaleron black hole. As for
the globally
charged black hole (monopole black hole), we can also
apply catastrophe
theory to the stability analysis, although the behavior of
the solutions is
more complicated.

KM would like to thank Gary W. Gibbons for discussion
about
the relationship between cusps and catastrophe theory, and
for
information about references.
We acknowledge L.D. Gunnarsen for his critical reading of
our
report.  This work was supported partially by the Grant-
in-Aid for
Scientific Research  Fund of the Ministry of Education,
Science and
Culture  (No. 04640312 and No. 0521801), by the Grant-in-
Aid for JSPS Fellows
(053769), by a Waseda
University Grant for Special Research Projects,  and by
The Sumitomo Foundation.

\newpage
\tiny

\hspace{-1.2cm}
\begin{tabular}{|l||l|l||l|l|l|l|}
\hline
\hline
& & & & & &\\
{\bf Black holes} &{\bf non-Abelian fields}
& {\bf Higgs fields}
& {\bf particles}& {\bf black hole modes}& {\bf mass
(entropy)}&$C_\#$\\[1em]
\hline
\hline
& & & & & &\\
1. colored BH&Yang-Mills field ($SU(2)$)&
{}~~~~~~~~~~~--- & BM particle
&1(unstable)
& $m<\infty$ & 2\\[.5em]
& $-\frac1{16\pi} {\rm Tr} F^2, ~~~F=dA +g A\wedge A$ & &
&  & &\\[1em]
\hline
& & & & & &\\
2. Skyrme BH&Skyrme field ($SU(2) \times SU(2)$)
& ~~~~~~~~~~~---& Skyrmion
&1(stable)
&finite (high)&0 \\[.5em]
& $-\frac1{32 g_S^2}{\rm Tr} F^2 - \frac14 f_S^2 {\rm Tr}
A^2, ~~~F=-dA$ & &BM type
&1(unstable)
&  finite (low) &1 or 3 \\[1em]
\hline
& & & & & &\\
3. Proca BH&  massive Yang-Mills (Proca) field
& ~~~~~~~~~~~---
& Procaon
&1(stable)
& finite (high) & 0 \\[.5em]
&$-\frac1{4\pi}[\frac14{\rm Tr} F^2 -\frac12 \mu^2 {\rm
Tr} A^2]$& & BM type
&1(unstable)
& finite (low)  & 1 or 3\\[1em]
\hline
& & & & & &\\
4. sphaleron BH&Yang-Mills field ($SU(2)$)& (complex
doublet)
& sphaleron
&1(``stable")
& finite (high) & 0 \\[.5em]
&$-\frac1{16\pi}{\rm Tr} F^2$ & $-\lambda (\Phi^{\dagger}
\Phi-\Phi_0^2)^2$& BM type
&1(unstable)  &  finite (low) & 1 or 3\\[1em]
\hline
& & & & & &\\
5. monopole BH&Yang-Mills field ($SU(2)$)& (real triplet)
& monopole
&1(stable)
& finite (high) & 0\\[.5em]
& $-\frac1{16\pi}{\rm Tr} F^2$ & $-\frac{\lambda}{4} (\
Phi^2-\Phi_0^2)^2$&
[BFM particle] & [1(unstable)]  & [finite (low)] & [0]\\
[1em]
\hline
\hline
& & & & & &\\
Reissner&electromagnetic field ($U(1)$)& ~~~~~~~~~~~---
&~~~~~~---&1(stable)
& $m<\infty$ &1\\[.5em]
{}~~-Nordstr\"{o}m BH&Yang-Mills field ($SU(2)$) & (real
triplet)
&~~~~~~---&1(stable or unstable)
& $m<\infty$ &1\\[1em]
\hline
\hline
\end{tabular}

\vspace{.3cm}

\footnotesize
\begin{center}
{\normalsize Table 1}
\end{center}
\begin{flushleft}
{\bf Table Caption}
\end{flushleft}
\baselineskip .65cm

\vskip 0.1cm
\noindent
\parbox[t]{2cm}{\bf Table 1:\\~}\ \
\parbox[t]{14.5cm}{The properties of 5 models including
non-Abelian fields.
 See text about the meaning of ``stable" for the sphaleron
black hole. BFM particle means one of two non-trivial
solutions
 found in \cite{BFM}, which
 is more massive than the usual monopole.  $C_\#$ denotes
how many times the sign of the
specific heat changes in the branch.  The Reissner-Nordstr
\"{o}m  black holes in
Einstein-Maxwell and Einstein-Yang-Mills-Higgs systems are
listed as references.
In order to define parameters in the theories such as
a gauge coupling constant $g$, we show
the Lagrangians of non-Abelian fields and the potentials
of Higgs fields.}\\[3em]

\begin{flushleft}
{\bf Figure Captions}
\end{flushleft}
\baselineskip .65cm

\vskip 0.1cm
\noindent
\parbox[t]{2cm}{\bf Figure 1:\\~}\ \
\parbox[t]{14.5cm}{The mass-horizon radius diagrams for
(a) the Skyrme black
hole with $f_S/m_P$ = (i) 0.01, (ii) 0.02, and (iii) 0.03,
(b) the Proca black
hole with $\mu/g m_P$= (i) 0.05, (ii) 0.10, and (iii)
0.15, and (c)
the sphaleron black
hole with  $\lambda=0.125$ and $\Phi_0/m_P$= (i) 0.1, (ii)
0.2, and (iii) 0.3.
$C$ is a cusp, where the black hole has
a maximum entropy. Beyond its entropy there is no non-
Abelian black hole. The
Schwarzschild black hole (the dot-dashed line) and the
colored black hole (the
dotted line) are also shown as references.}\\[1em]
\noindent
\parbox[t]{2cm}{\bf Figure 2:\\~\\~}\ \  \
parbox[t]{14.5cm}{The solution curve in
the three dimensional space of ($M,B_H, S$) and its
projections onto
 each two dimensional planes for
a Skyrme black hole with $f_S/m_P = 0.02$. The cusp $C$ in
$M$-$S$ plane is a
critical point for stability.
For the fixed control parameter $M$, two solutions are at
extremal
points on the Whitney surface; the maximal one is stable,
while the
minimal one is unstable.  Beyond the critical point $C$,
there is no
extremal point, i.e.,  no non-Abelian black hole.}

\begin{thebibliography}{99}
\bibitem{note1} In what follows, we include the non-
trivial particle solutions
                in a family of the non-Abelian black holes
as the limit of
                 vanishing horizon radius.
\bibitem{note2}  As Aichelburg and  Bizon claimed\
cite{AB}, the mass is not
                 always a good indicater to analyze the
stability.
\end{thebibliography}
\end{document}